%% file: main.tex
\documentclass[aps, prd, a4paper, 10pt, showpacs, superscriptaddress, floatfix, nofootinbib, reprint, longbibliography]{revtex4-1}

\usepackage{geometry}
\usepackage[activate={true,nocompatibility},final,tracking=true,kerning=true,factor=1500,stretch=10,shrink=10]{microtype}
\usepackage{scalerel}[2016/12/29]
\usepackage{bm, amssymb, amsmath, amsfonts, multirow, float, siunitx, xfrac, url, lipsum}
\usepackage{units, verbatim, multirow, booktabs}

\usepackage[usenames]{color}
\usepackage{graphicx, subfigure}

\usepackage{hyperref}
\hypersetup{
	colorlinks=true, 	
	linkcolor=cyan,		
    citecolor=red,		
}

\newcommand{\IITGn}{Indian Institute of Technology Gandhinagar, Gujarat 382355, India}
\newcommand{\CMI}{Chennai Mathematical Institute, Plot H1, SIPCOT IT Park, Siruseri, 603103 Tamilnadu, India.}
\newcommand{\PennState}{Institute for Gravitation \& the Cosmos, Physics Department, Penn State, PA 16802, USA.}

\newcommand\Hyp{\scaleobj{0.9}{\mathcal{H}}}
\newcommand\inv[1]{#1\raisebox{1.15ex}{$\scriptscriptstyle-\!1$}}
\newcommand\ginv[1]{#1\raisebox{1.15ex}{$\scriptscriptstyle-\!3$}}

\newcommand{\coloneq}{\mathrel{\resizebox{\widthof{$\mathord{=}$}}{\height}{ $\!\!\resizebox{1.2\width}{0.8\height}{\raisebox{0.23ex}{$\mathop{:}$}}\!\!=\!\!$ }}}
\newcommand{\norm}[1]{\left\lVert#1\right\rVert}

\DeclareMathOperator*{\argmax}{arg\,max}

\DeclareMathOperator{\Expect}{\mathbb{E}}

\def \msun	{{M}_\odot}

\def \seobnrhm {\text{SEOBNRv4HM}}

\def \myTitle {Unveiling the spectrum of inspiralling binary black holes}

\begin{document}

\title{\myTitle}

\author{Soumen~Roy}
\email{soumen.roy@iitgn.ac.in}
\affiliation{\IITGn}

\author{Anand~S.~Sengupta}
\email{asengupta@iitgn.ac.in}
\affiliation{\IITGn}

\author{K.~G.~Arun}
\email{kgarun@cmi.ac.in}
\affiliation{\CMI}
\affiliation{\PennState}




\begin{abstract}

The higher-multipoles of gravitational wave signals from coalescing compact binaries play a vital role in the accurate reconstruction of source properties, bringing about a deeper and nuanced understanding 
of fundamental physics and astrophysics. Their effect is most pronounced in systems with asymmetric masses having an orbital geometry that is not face-on.
The detection of higher-multipoles of GW signals from any single, isolated merger event is challenging, as there is much less power in comparison to the dominant quadrupole mode.
In this paper, we present a new method for their detection by combining multiple events observed in interferometric gravitational wave detectors. Sub-dominant modes present in (the inspiral part of) the signal 
from separate events are stacked using time-frequency spectrogram of the data. We demonstrate that this procedure enhances the signal-to-noise ratio of the higher-multipole components and thereby leads to increased chances of their detection. 
%
From Monte-Carlo simulations we estimate that a combination of $\sim 100$ events observed in two-detector coincidence can  lead to the detection of the higher-multipole components with a $\geq$ 95\% detection probability. 
The advanced-LIGO detectors are expected to record these many binary black hole merger events within a month of operation at design sensitivity.
We also present results from the analysis of data from O1 and O2 science runs containing previously detected events using our new method.

\end{abstract}

\preprint{{\color{red}[LIGO-P1700019]}}

\maketitle

\section{Introduction}
\label{sec:intro}
The first detection of gravitational wave (GW) from a merging
binary black hole (BBH)~\cite{gw150914} has ushered in a new era in observational astronomy and fundamental
physics. 
From current estimates of the rate of BBH mergers, one expects future gravitational wave detectors to observe a large number of events which can reveal the diversity in population of compact binaries.  
Among compact binaries with precession and
orbital eccentricity, an important class of sources that has eluded us
thus far are the ones which show signatures of higher-harmonics in the 
gravitational wave signal.
%

According to general relativity (GR), inspiralling compact binaries emit gravitational waves predominantly at twice the orbital frequency. 
In addition, the signal contains higher-harmonics at other integer multiples of this fundamental frequency, but whose amplitudes are suppressed in comparison to the dominant quadrupole mode~\cite{Arun:2004ff,VanDenBroeck:2006qu, Broeck_2007}.
%
%
%
Their relative strength 
also depend on the orientation of the binary with respect to the observer's line of sight (zero for
``face-on" binaries) and the mass ratio of the binary constituents (odd multipoles are zero
for equal mass systems). 

While it is difficult to detect the faint higher-multipoles of the signal, their subtle interplay with the dominant mode adds to the overall complexity and richness of the signal, and remarkably improves the accuracy of estimated source parameters. As such, higher-multipoles present in the signal can 
%
pave the way for new tests of GR~\cite{Dhanpal:2018ufk}, resolve the two states of gravitational wave polarization~\cite{Jennrich:1997if}, measure the inclination angle~\cite{Arun:2014ysa} from neutron star - black hole compact binary systems and thereby constrain possible jets~\cite{Arun:2014ysa, lIGO_Fermi_2017}.
%

The present generation of interferometric GW observatories are biased towards detecting comparable-mass inspiraling binaries 
in the face-on or face-off orientation to the line of sight. As such, they are unlikely to detect higher-order modes from a single observation. However, a combination of several observations could unravel these weak signals as shown here.

Earlier studies have capitalized on the constant frequency of the final black hole's ringdown modes, and developed algorithms to stack the {\emph post-merger} ringdown signals.
%
%
These include time-domain coherent mode stacking~\cite{ModeStack-2017} and in the time-frequency domain~\cite{OBrien:2019hcj}. 
A recent study has identified one overtone of the dominant ringdown mode~\cite{Isi-2019, Giesler-2019}, using time-domain multimode analysis~\cite{Carullo-2019}. 
Tests of GR with higher-order modes of ringdown signals from multiple BBH observations have also been posited~\cite{Brito-2018, Meidam-2014} using  Bayesian model selection methods.


We are unaware of any work in literature that deals with the problem of combining inspiral-meger parts of GW signals -- possibly due to the difficulty posed by their time-varying instantaneous frequency, especially in the late-inspiral stages. We address this problem in this paper by presenting a new method that simultaneously stacks all the multipoles present in these signals from independent events.


\vspace{0.5\baselineskip}
 

%
\section{Data and Signal}
The GW wave signal $h(t)$ 
propagating along an arbitrary direction $(\iota, \phi_0)$ in the source frame, can be decomposed over the 
spin-weighted spherical harmonic basis (with spin-weight $-2$) as:
\begin{equation}
\label{eq:HOM}
  h(t; \iota, \phi_0, \vec{\lambda} ) = \sum_{\ell=2}^{\infty}\sum_{m=-\ell}^{\ell} {}_{-2}Y^{\ell m}(\iota, \phi_0) \, h_{\ell m}(t; \vec{\lambda}), 
\end{equation}
where, $h_{\ell m}(t; \vec{\lambda}) = A_{\ell m}(t; \vec{\lambda}) \; e^{i \Phi_{\ell m}(t; \vec{\lambda})}$ represents the $(\ell, m)$ mode of the signal described by the corresponding amplitude $A_{\ell m}(t; \vec{\lambda} )$ and phase $\Phi_{\ell m}(t; \vec{\lambda})$; and where $\vec{\lambda}$ 
represents the set of intrinsic parameters. 
In particular, for non-precessing spinning BHs, the inspiral 
phase of an arbitrary $(\ell, m)$ mode can be expressed in terms of the phase of the $(2,2)$ mode alone: ${ \Phi_{\ell m}(t; \vec{\lambda}) \simeq  (\sfrac{m}{2}) \; \Phi_{22}(t; \vec{\lambda}) }$.
%
%
This translates to a relation between their instantaneous frequencies:  ${ f_{\ell m}(t; \vec{\lambda}) = \dot{\Phi}_{\ell m} \simeq (\sfrac{m}{2}) \; f_{22}(t; \vec{\lambda}) }$ - which can be used to define an arbitrary time-frequency `track' scaled with respect to the trajectory of the $(2,2)$ track, 
%
%
\begin{equation}
\label{eq:tftrack}
f_{\alpha}(t; \vec{\lambda}) = \alpha \, f_{22}(t; \vec{\lambda}), 
\end{equation}
where $\alpha > 0$ is a scaling factor. 
The specific tracks of the $(\ell, \pm m)$ harmonic of the signal are obtained by setting ${\alpha=\sfrac{m}{2}}$ in Eq.~\eqref{eq:tftrack}.
%

The relationship between the phase of the harmonics 
of a GW signal is valid over the inspiral and merger regime, and is vital to the method presented in this paper. 
Using a time-frequency spectrogram of the signal, this relation is leveraged for accumulating the signal energy along tracks parametrized by the scaling parameter $\alpha$, thereby decoupling the different modes of the GW signal. 
%
Note that while \emph{all} the $(\ell \geq m, \: \pm m)$ modes of the signal follow the same track for $\alpha = \sfrac{m}{2}$, the energy along such a track is dominated by the $(\ell=m, \: \pm m)$ mode.
%

The time-frequency representation of any time-series $x(t)$ is obtained from its scaleogram $\tilde{X}(\tau, f)$ defined to be the absolute square of its continuous wavelet  transformation (CWT)  
%
calculated in the Gabor-Morlet~\cite{Morlet-1984} wavelet basis (see Appendix~\ref{sec:CWT}). The latter is characterised by the time-translation ($\tau$), scale ($a$) and central frequency ($f_0$) parameters.\\
The energy $\tilde{x}(\tau, f)$ contained in a specific pixel centred on $(\tau, f)$ can be obtained from the scaleogram:
\begin{equation}
\label{eq:CWTPixleEnergy}
  \tilde{x}(\tau, f) \equiv \frac{1}{C_g} \tilde{X}(\tau, a) \; \frac{\Delta a}{a^2} \Delta \tau,
\end{equation}
where, $\Delta \tau$ and $\Delta a$ are the time and scale spacings respectively, and $C_g$ is the admissibility constant. \\
$f_0$ regulates the spectral leakage of the signal over the $\tau - a$ plane, and was optimally chosen to maximise the energy in pixels along the $f_{22}(t)$ trajectory.

%
%

We adopt the following notation: the {\emph{whitened}} ``on-source" detector data time-series encompassing the event epoch 
is denoted by $y(t) = n(t) + s(t; \vec{\lambda})$: consisting of `ideal' detector noise $n(t)$ having a normal distribution $\mathcal{N}(0,1)$; and an embedded gravitational wave signal $s(t; \vec{\lambda})$
Their corresponding spectrograms, calculated using Eq.~\eqref{eq:CWTPixleEnergy} are denoted by $\tilde{y}$, $\tilde{n}$ and $\tilde{s}$ respectively. 
The \texttt{aLIGO} power spectral density~\cite{aLIGO_ZDHP} is used to whiten the data and signals unless stated otherwise.

The embedded signal $s(t;\vec{\lambda})$ is constructed 
from theoretical waveform models which include higher-order modes. The signal's intrinsic parameters $\vec{\lambda}$  is determined from the measurement of the dominant $(2,2)$ quadrupole mode. Data samples that lie few tens of seconds away from the detection epoch (i.e. off-source data segments) are assumed to contain no astrophysical GW signal, and provide representative samples of the noise $n(t)$. 

The template vector $S(\alpha) \in \mathbb{R}^d$ is calculated from 
$\tilde{s}(\tau, f)$; by summing over the pixels along time-frequency arcs given by Eq.~\eqref{eq:tftrack}:
\begin{equation}
\label{eq:sigmodel}
  S(\alpha) = \sum_{\tau=t_c - \Delta \tau}^{t_c} \tilde{s} \left ( \tau, \, f = \alpha f_{22}(\tau; \vec{\lambda}) \right ),
\end{equation}
leading up to the epoch $t_c$ at which the orbiting masses reach the innermost stable circular orbit (ISCO). 
The scaling parameter $\alpha$ takes $d$-discrete steps in the interval $[\alpha_\text{min}, \:\alpha_\text{max}]$ .

%
In practice, we curtail the summation at 
an epoch when the GW frequency reaches $0.6\: f_{\text{ISCO}}$ to avoid $\gtrsim 1\%$ overlap of power between multipoles caused due to the finite resolution of spectrograms. We choose $\Delta \tau = 0.5\,\si{\second}$ to focus on the late-inspiral stage where the signal amplitude is relatively higher.

%

The data vector $Y(\alpha)$ is constructed from `on-source' data by substituting $\tilde{s}$ on the RHS of Eq.~\eqref{eq:sigmodel} with $\tilde{y}$. In a similar manner, spectrograms $\tilde{n}$ of off-source data segments provide an ensemble of noise vectors $N(\alpha)$.

\begin{figure}
\centering
  \includegraphics[width=0.495\textwidth]{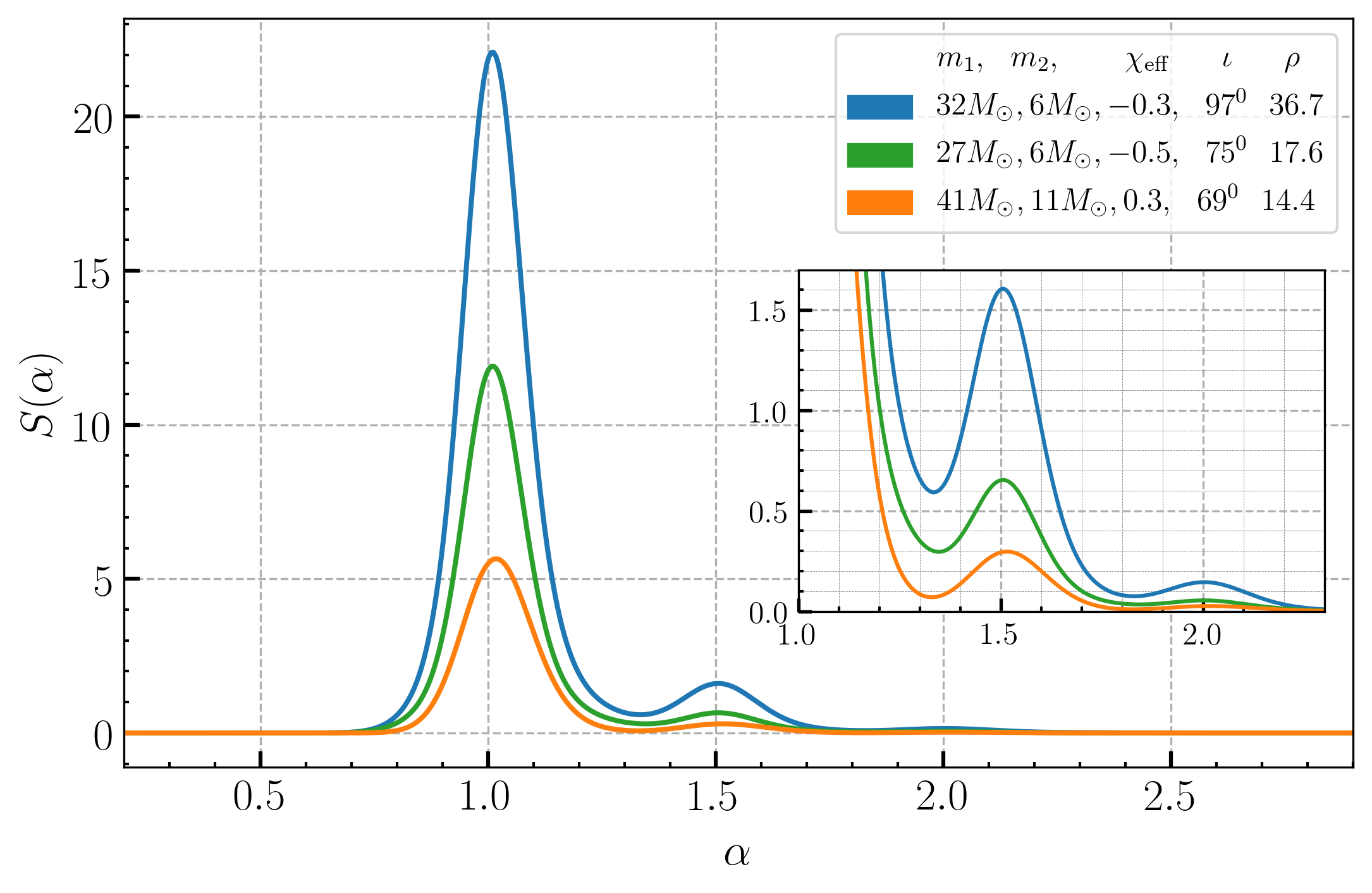}
  \caption{Template vectors $S(\alpha)$ for three different non-precessing asymmetric BBH systems generated using the 
  $\mathrm{SEOBNRv4HM}$ waveform model~\cite{Cotesta-2018} consisting of $(2, 1)$, $(2, 2)$ $(3, 3)$, $(4, 4)$ and $(5, 5)$ 
  multipoles. Peaks at $\alpha = 1.0$  and $\alpha=1.5$ indicate the relative energy of the $m=2$ and $m=3$ 
  modes. 
  }
  \label{fig:sigmodel}
\end{figure}

We illustrate $S(\alpha)$ vectors 
for three non-precessing BBH systems in Fig.~\ref{fig:sigmodel}. A dominant peak at $\alpha=1$ corresponds to the quadrupole mode and a prominent peak at $\alpha=\sfrac{3}{2}$ is observed for all three systems corresponding to the energy present in the 
next-highest $(3,3)$ mode of the signal. In contrast, the peaks at $\alpha = \sfrac{5}{2}, \sfrac{4}{2}$ and $\sfrac{1}{2}$ are much smaller, in proportion to the relative energy in these modes. The height of these peaks depend on the signal parameters and sensitivity of the detectors
%
%
whereas the peak-widths result from the finite time-frequency resolution. 

For a hypothetical spectrogram having an arbitrarily fine pixel resolution, the $S(\alpha)$ vector will be a sum of several Dirac-$\delta$ functions located at $\alpha=\{0.5, 1.0, 1.5, 2.0, 2.5\}$.
The peaks of $S(\alpha)$ from all the three events having different parameters occur at the same value of $\alpha$; indicating the possibility of simultaneously stacking the quadrupole and other subdominant modes of several events over the $\alpha$ parameter, thereby enhancing their detectability.  

\vspace{0.5\baselineskip}

\section{Single event detection statistic}

Assuming an unambiguous detection of the dominant quadrupole mode of a BBH merger signals in aLIGO-like detectors, made by standard data-analysis pipelines, we now outline a follow-up statistical test for the detection of their next-loudest $(3, 3)$ modes. The method presented here can be extended to other multipoles.

We propose the following three composite hypotheses:
%
\begin{equation}
\label{eq:DefHypotheses}
  \begin{split}
    \Hyp_0 : Y(\alpha) &= N(\alpha),  \\
    \Hyp_2 : Y(\alpha) &= N(\alpha) + a_2 \, S_2(\alpha),  \\
    \Hyp_3 : Y(\alpha) &= N(\alpha) + a_2 \, S_2(\alpha) + a_3 \, S_3(\alpha),
  \end{split}
\end{equation} 
where $N(\alpha)$ is the contribution from random instrumental noise in the data and where $S_2(\alpha)$ and $S_3(\alpha)$ are the contributions from the $m=2$ and $m=3$ multipoles of the best-fit embedded signal. 
The signal amplitude depends on the extrinsic parameters of the signal 
that are not well estimated from the 
dominant quadrupole mode of the signal. This uncertainty is incorporated 
through the free overall amplitude parameters $a_2$ and $a_3$ whose numerical values are simultaneously determined by maximizing the logarithmic likelihood ratio (LLR) $\Lambda_3(a_2, a_3)$ of observing $Y(\alpha)$ under $\Hyp_3$ as compared to the null hypothesis $\Hyp_0$:
%

%
\begin{equation}
\label{eq:Amplitudefactor}
a_{2, 3}^{\ast} = \argmax_{a_2,\, a_3}\, \Lambda_3(a_2, a_3).
\end{equation}
The evaluation of $\Lambda_3$ assumes that each of the  $N(\alpha)$ noise vectors 
is a correlated $d$-dimensional Gaussian random variable. Their correlation is captured by the covariance matrix which can be calculated numerically from the ensemble average of several noise vectors, along with their ensemble average $\mu(\alpha)$ 
(see Appendix~\ref{sec:EstimateNoise}). 
%

We define a new detection statistic $\beta$ by subtracting the contribution of the $m=2$ multipole in $Y(\alpha)$ so as to measure the contribution from only the $m=3$ multipole of the signal (see Appendix~\ref{sec:SupHypothesisTesting}):
%
\begin{equation}
  \label{eq:beta_defn}
  \beta = \langle Y(\alpha) - \mu(\alpha) - a_2^{\ast} \, S_2(\alpha) \mid a_3^{\ast} \, S_3(\alpha) \rangle  /\gamma_3,
\end{equation}
 where, $\gamma_3 = \lVert a_3^{\ast} \, S_3(\alpha)\rVert$ is the maximised template norm. 
 Here $\langle \cdot \vert \cdot \rangle$ denotes the covariance matrix weighted inner-product between two vectors.

%
Cross-terms between the embedded signal and noise in the spectrogram $\tilde{y}$ of the on-source data segment increases the variance of the background distribution $p \left ( \beta \mid \Hyp_2 \right )$. The variance also depends on the strength of the embedded signal. In the absence of cross-terms (or a weak signal), $p (\beta \mid \Hyp_2) \sim \mathcal{N}(0, 1)$.

When comparing detection statistic for different independent events and also where multiple events are combined, we scale $\beta$ by the standard deviation of the corresponding background distribution. This ensures that all the events have  $\mathcal{N}(0, 1)$ background distributions, making meaningful comparisons of the detection statistic possible.  
The nominal detection threshold for the $\mathcal{N}(0, 1)$ background distribution can be set at $\beta^\ast = 2.325$ corresponding to a fixed false-alarm probability of $1\%$, 
%

%


\vspace{0.5\baselineskip}


%
\section{Stacking up multiple BBH observations}
At design sensitivity, the advanced LIGO/Virgo detectors are expected to observe signals from several tens of coalescing binary blackholes every week. We now show how data from these observations can be combined (or stacked) to enhance the signature of higher-multipole signal components. We can also stack data from different detectors for the same observation, by treating them as independent events. The ``combined detection statistic" $\beta$ is also given by Eq.~\eqref{eq:beta_defn} where, one uses the {\emph{stacked versions}} of various pieces that appear on the RHS.
 
The combined $Y(\alpha)$ vector is constructed by adding the on-source $Y^{(j)}(\alpha)$ for each of the $j = 1, 2, \cdots, n_0$ observations: $Y(\alpha) = \sum_{j} Y^{(j)}(\alpha)$, 

The combined template vectors 
are constructed 
by adding the single-event template vectors: $a^{\ast}_{2,3} \, S_{2,3}(\alpha) = \sum_{j} a_{2,3}^{\ast(j)} \, S_{2,3}^{(j)}(\alpha)$. It is implied that the maximised amplitude coefficients $a_{2,3}^{\ast(j)}$ are obtained from Eq.~\eqref{eq:Amplitudefactor}, separately for each event.

%
The ensemble of noise vectors 
from off-source data segments around the $j\textsuperscript{th}$ event are also similarly combined. 

Finally, they are plugged into Eq.~\eqref{eq:beta_defn} to calculate the detection statistic.

In Fig.~\ref{fig:increase_SNR} we show that the average detection statistic $\langle \beta \rangle \propto \sqrt{n_0}$ when $n_0$ {\emph{identical}} events are combined using the method presented here (where the average is obtained over injections made in many noise realizations). From this scaling, we establish the fully coherent nature of stacking the higher-multipoles modes.  
In contrast, combining the events in a Bayesian model selection study through the product of the Bayes factors of the events leads to a $\sim n_0^{\sfrac{1}{4}}$ scaling of the SNR~\cite{ModeStack-2017}.
It also turns out that only those events with ``comparable'' signal norms are worth stacking. The explanation for this fact, leading to a prescription for choosing the useful events is available in the text around Eq.~\eqref{eq:selectEvents}.

\subsection{Demonstrating the coherent nature of stacking}
\label{sec:CoherentSum}


%
Let us assume that we have a set of identical injections containing the dominant ($m=2$) and next-higher ($m=3$) harmonics of the signal in $n_0$ realisations of \texttt{aLIGO} noise. Let the strength of the injected $m=3$ component be such that the norm of its signal vector $\norm{S_3(\alpha)} = \gamma_3^{\text{inj}}$.

As discussed earlier, Eq.~\eqref{eq:beta_defn} gives the single-event detection statistic and measures the strength of the $m=3$ multipole of the signal. The same expression can be used for the combined detection statistic (after stacking multiple events), except that the pieces in the RHS of this equation must now be replaced by their stacked counterparts. 
In the present case, these pieces (after stacking) are given by:
%
\begin{align}
  Y(\alpha)			&= \sum_{j=1}^{n_0} Y^{(j)}(\alpha), \label{eq:combdataSup}, \\
  \mu(\alpha) 		&= \Expect \left[ \sum_{j=1}^{n_0} N^{(j)}(\alpha) \right], \\
  a_{2}^{\ast} \, S_{2}(\alpha)	&= \sum_{j=1}^{n_0} a_{2}^{\ast (j)} \, S_{2}^j(\alpha) \\
  %
  a_{3}^{\ast} \, S_{3}(\alpha)	&= \sum_{j=1}^{n_0} a_{3}^{\ast (j)} \, S_{3}^j(\alpha) \label{eq:combsignormSup1}
  %
 \end{align}
 and finally, using the fact the combined noise variance matrix is given by $\Sigma = n_0 \, \Sigma^j$, the norm of the combined template can be shown to be
 \begin{align}
  \gamma_3 		&= \norm{a_{3}^{\ast} \, S_{3}(\alpha)} = a_{3}^{\ast}\, \gamma_3^{\text{inj}}\; / \sqrt{n_0}.
  \label{eq:combsignormSup}
\end{align}

%

The combined data vector in Equation~\eqref{eq:combdataSup} can be further expanded as:
\begin{align}
  Y(\alpha)  &=	\sum_{j=1}^{n_0} \, N^{j}(\alpha) + S^j_2(\alpha) + S^j_3(\alpha) \nonumber \\
             &	\qquad\quad + {\sf X}^j_2 + {\sf X}^j_3 + {\sf X}^j_{23} \label{eq:CombDataSup}
\end{align}
where ${\sf X}^j_{2}$ (${\sf X}^j_{3}$) denote cross-terms between noise and $m=2$ ($3$) multipoles of the signal in the spectrogram of the on-source data from the {$j$-th} event, while ${\sf X}^j_{23}$ denotes the cross-term between these two multipoles. Plugging this in Eq. S15, rearranging and noting that ${\sf X}^j_{23} = 0$, we have 

\begin{align}
  \langle \beta \rangle  & =  \left \langle \sum_{j=1}^{n_0} N^{j}(\alpha) - \mu(\alpha) + {\sf X}^j_{2} + {\sf X}^j_{3} \mid a_3^{\ast} S_3(\alpha) \right \rangle / \gamma_3 \nonumber \\
  & \quad + \left\langle n_0 \, S_2(\alpha) - a_2^{\ast} \, S_2(\alpha) \mid a_3^{\ast} \, S_3(\alpha) \right\rangle / \gamma_3 \nonumber\\
  & \quad + \langle n_0 \, S_3(\alpha) \mid a_3^{\ast} \, S_3(\alpha) \rangle / \gamma_3 \nonumber \\
  & =  n_0 \, a_3^\ast \norm{S_3(\alpha)}^2 /n_0 \gamma_3 \nonumber \\
  & = \sqrt{n_0} \,  \gamma_3^{\text{inj}}.
\end{align}
Recalling that the mean of the single-event detection statistic is equal to the signal norm $\gamma_j = \gamma_3^{\text{inj}}$, we have $\langle \beta \rangle / \langle \beta_j \rangle = \sqrt{n_0}$.
 
\begin{figure}[t]
  \centering
  \includegraphics[width=0.49\textwidth]{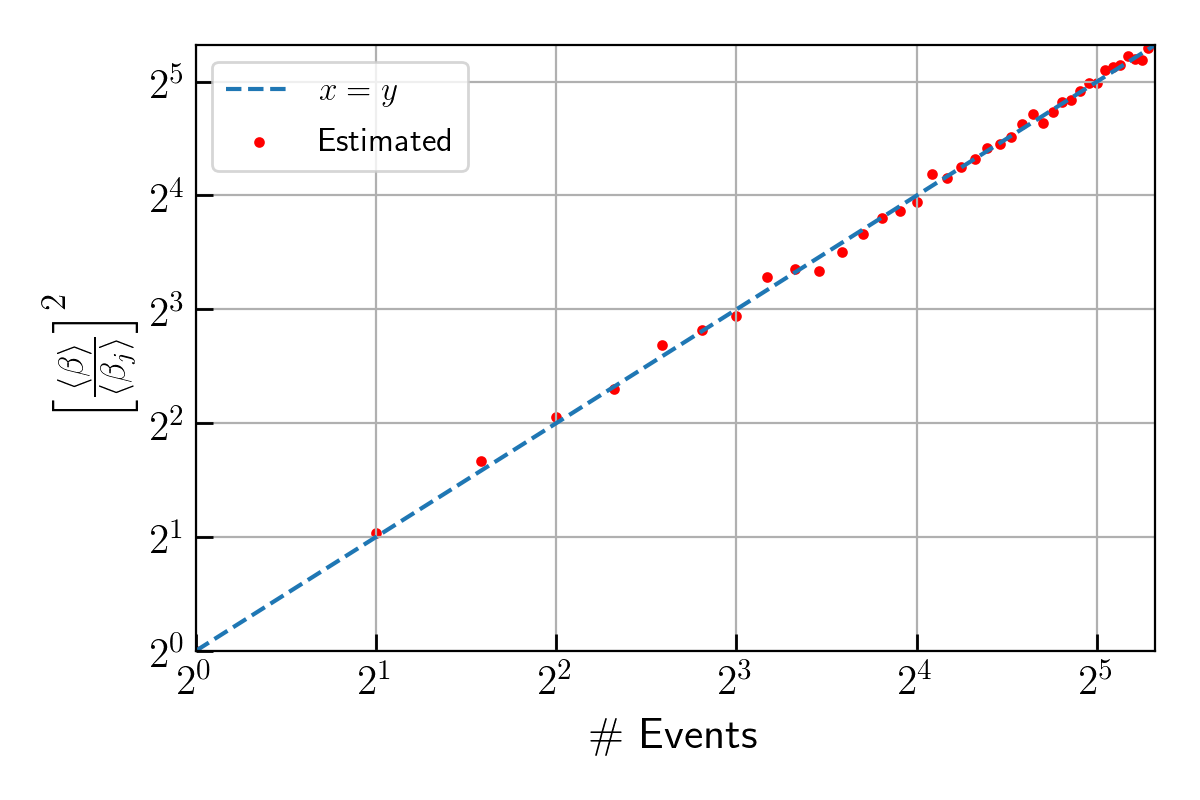}
  \caption{The plot of $\left[ \langle \beta\rangle/\langle\beta_j\rangle\right]^2$ versus number of events follows a straight line with unit slope, where all the events are identical. Ensemble averages are taken over $300$ realisations of ideal \texttt{aLIGO} noise. The plot implies that the stacking algorithm is coherent where with the average detection static (after stacking $n_0$ identical events) scales as $\sqrt{n_0}$.}
  \label{fig:increase_SNR}
\end{figure}

In Fig.~\ref{fig:increase_SNR}, we stack a number of identical events (embedded in ideal Gaussian noise) and compare the ratio $\left[ \langle \beta\rangle/\langle\beta_j\rangle\right]^2$ with the analytical result obtained above. The agreement between the two shows that the stacking method presented in this paper indeed combines the events {\it coherently} with an increase of the statistic by a factor of $\sqrt{n_0}$.


A subtle point in combining the events can be illustrated by considering only two events with identical intrinsic parameters, with indices $j=1,2$ such that their observed norms are in the order  $\gamma_3^{(1)} > \gamma_3^{(2)}$. 
%
The combined template norm is {$\gamma_3 = \gamma_3^{(1)} ( 1 + \gamma_3^{(2)} / \gamma_3^{(1)})/\sqrt{2}$}.
Obviously, the combined $\gamma$ exceeds $\gamma_3^{(1)}$ only when 
$\gamma_3^{(1)}/\gamma_3^{(2)} \geq (\sqrt{2} - 1)$. 
This can be generalized for ${n_0}$ events assumed to be first arranged in a descending order of their norms 
such that $\gamma_3^{(1)} > \gamma_3^{(2)} > \cdots  > \gamma_3^{(n_0)}$. One chooses to `\emph{optimally}'  combine a subset of $n^\prime_0 \leq n_0$ events where: 
\begin{equation}
\label{eq:selectEvents}
  n'_0 = \argmax_{j \leq n_0 }\left \{ \left(\sum_{i=1}^j \gamma_3^{(i)}\right)^2 / j \right \}.
\end{equation} 
This leads to the maximum possible $\langle \beta \rangle$ after stacking. Thus, only those events can be combined whose signal norms are 'comparable' as argued above.

\section{Prospects in Advanced LIGO}
We present the results of a Monte-Carlo simulation using a set of $2500$ aligned-spin, non-precessing BBH systems having optimal quadrupole-mode SNR ${\rho_{22} \geq 8}$, to quantify the chances of observing the higher-multipoles in \texttt{aLIGO}-like detectors. 


The sources were drawn from an astrophysical population assuming a uniform merger rate density of $53 \,\ginv{\text{Gpc}}\inv{\text{yr}}$ in the co-moving volume for stellar-mass black holes, inferred from \texttt{aLIGO}'s \texttt{O1} and \texttt{O2} science runs~\cite{O1O1RatesAndPop}. These events are expected to be detected by current data analysis pipelines in \texttt{aLIGO} data within $\sim 1.5$ years of observation at design-sensitivity. 

%
%
\begin{figure}[t]
  \includegraphics[width=0.495\textwidth]{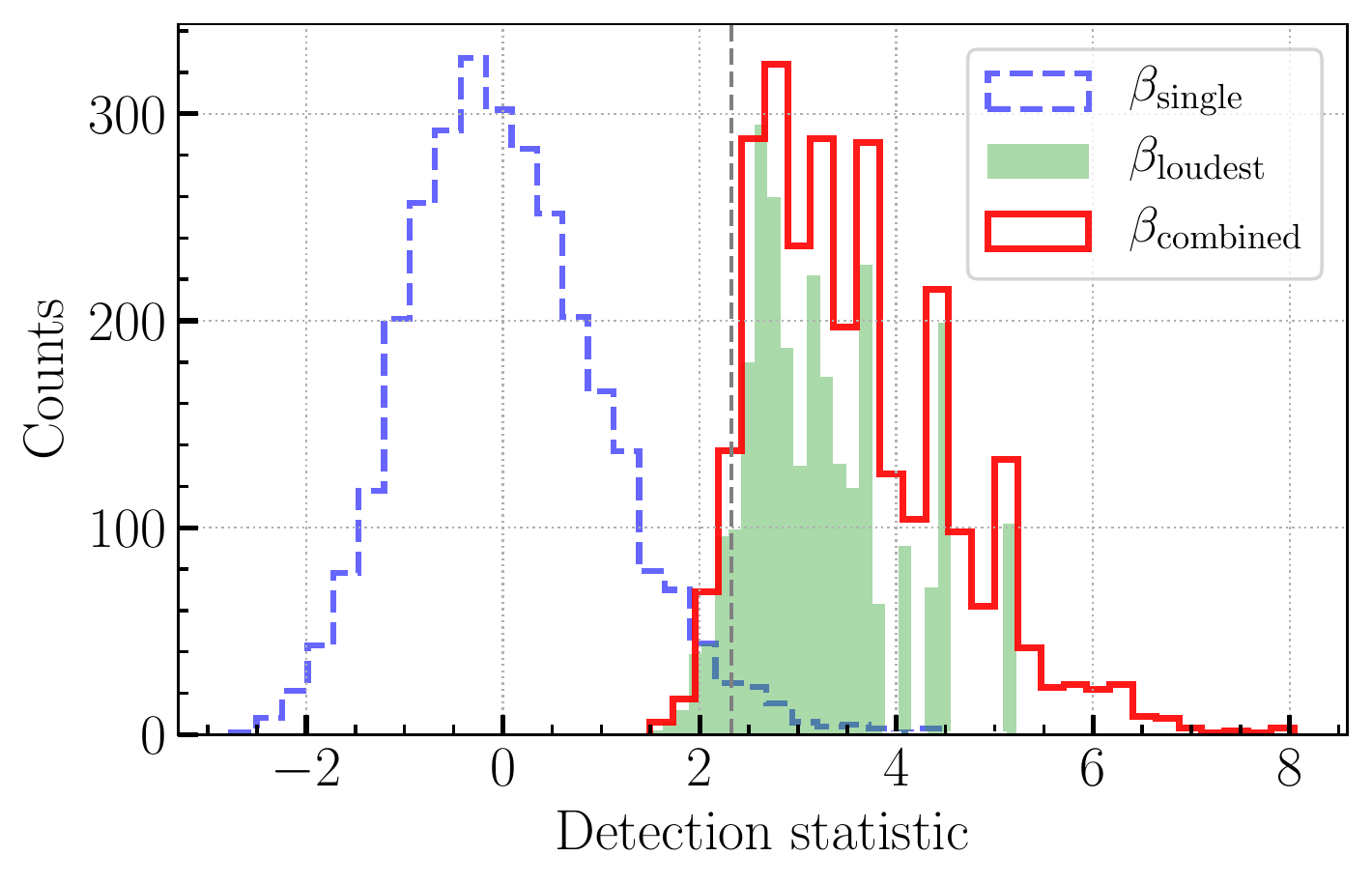}
  \caption{Distributions of the single-event detection statistic $\beta_{j}$ (dashed-blue trace) from the entire set of simulated events observed in  a \texttt{aLIGO} detector and of the combined statistic $\beta$ (in red), after optimally stacking a subset of $n_0 = 100$ randomly chosen events. This should be contrasted with the filled green histogram showing the detection probability obtained from choosing the maximum $\beta_j$ out of the same subset of 100 events without stacking.
There is a 95\% probability of detecting higher-multipoles after stacking 100 events. 
  }
  \label{fig:multiple_evn_LLR}
\end{figure}

The component masses (in $\msun$ units) were chosen between ${5 \leq m_{1,2} \leq 50}$ 
with the primary mass $m_1$ from  $p(m_1) \propto m_1^{-2.3}$ and 
$m_2$ from a uniform distribution $p(m_2) \sim \mathcal{U}[5,\: m_1]$. 
The dimensionless spins were drawn from $\mathcal{U}[-1,\: 1]$.
The sources were uniformly distributed over the celestial sphere up to a redshift of  $z = 1.4$, and their inclination angle 
isotropically distributed.
Redshift-luminosity distance conversions were made assuming the $\Lambda\text{CDM}$ cosmological model~\cite{Planck2015}. 
GW signals including sub-dominant modes were generated using the $\seobnrhm$ waveform model for each of the playground events and injected in synthetic Gaussian noise 
to mimic \texttt{aLIGO} data. Thereafter, single ($\beta_j$) and combined ($\beta$) detection statistic were calculated.

Fig.~\ref{fig:multiple_evn_LLR} shows the distribution of the single-event detection statistic $p(\beta_j \vert \Hyp_3)$ obtained from all the events in the playground set. 
By integrating the distribution above the detection threshold ${\beta^\ast}$, 
we find that the probability of detecting higher-multipoles from single events 
is only $3\%$.

%
Next, several subsets of $n_0 = 100$ events were chosen at random from the playground set through a bootstrapping procedure, and stacked 
using the prescription in Eq.~\eqref{eq:selectEvents}. 
For the same subset of events we also calculate $\beta_{\text{loudest}} = \max_j \beta_j \ j=1,\cdots,n_0$, the loudest single-event statistic without stacking.

Integrating over the distribution of the combined detection statistic above ${\beta^\ast}$: 
we find that stacking $100$ events leads to 
detection probability of 95\%. 

\begin{figure}[t]
  \includegraphics[width=0.495\textwidth]{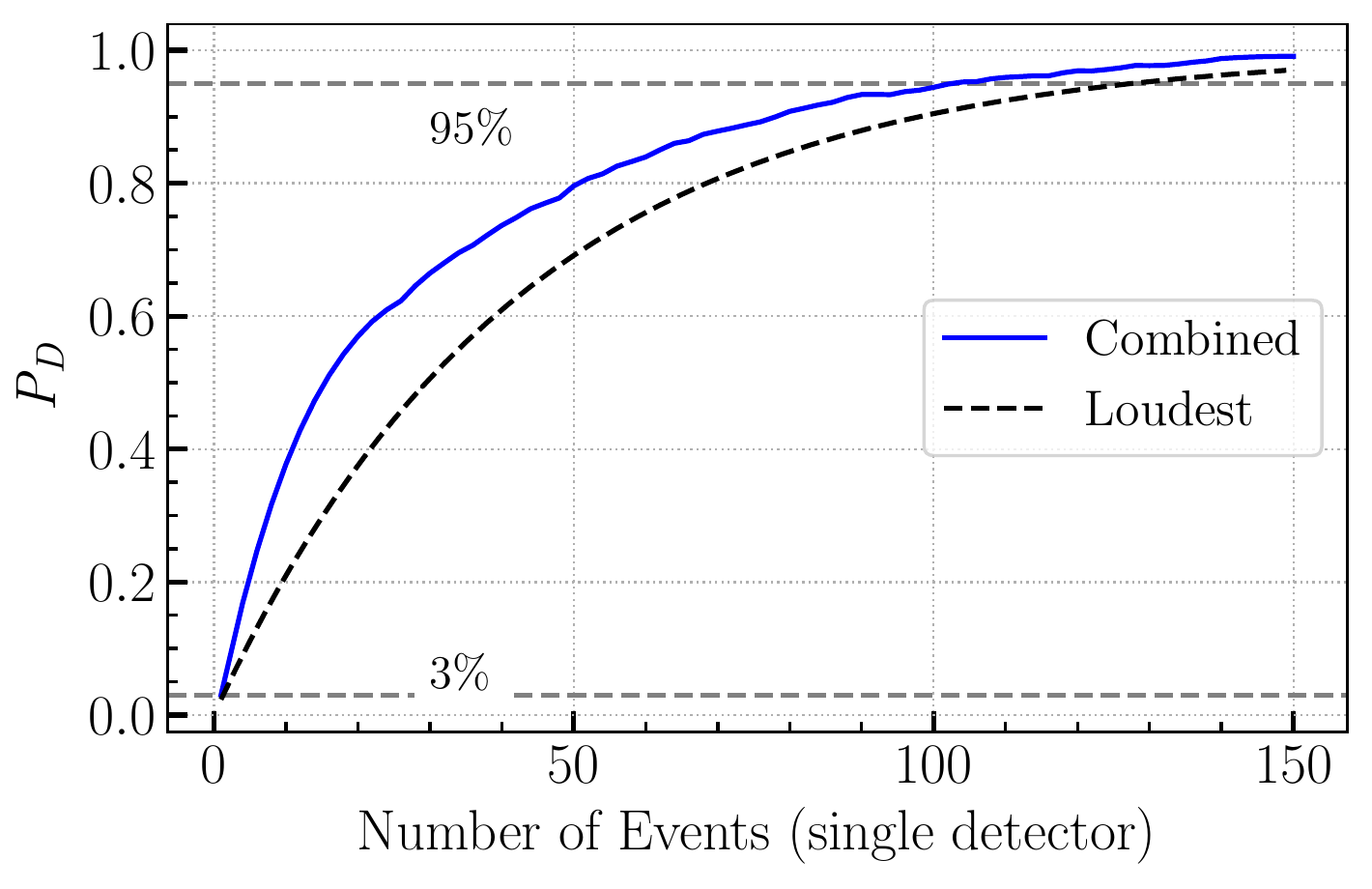}
  \caption{Plot showing the increase in detection probability ($P_D$) of higher-multipoles as more events are stacked using the method outlined in the text. As can be seen, stacking $n_0$ events is more efficient in comparison to choosing the loudest single event statistic. From the plot, we find that a nominal value of $P_D=95\%$ is reached after stacking 100 events. In comparison, the same target is reached in the latter case by considering 130 events. 
  }
  \label{fig:probd}
\end{figure}

In Fig.~\ref{fig:probd}, we quantify the detection probability $P_D$ by varying the number of stacked/combined events $n_0$ 
As expected, $P_D$ (calculated at 1\% false-alarm) grows monotonically with the number of stacked events, reaching 95\% for $100$ events and 99\% for 145 stacked events, respectively. 
In contrast, the same detection probability is achieved using the $\beta_{\text{loudest}}$ statistic from $n_0 = 130$ events. This shows the advantage of stacking events for detecting higher-order modes. This is particularly true for a hypothetical scenario where all the single events are below the threshold of detection, i,e, $\beta_j < \beta^\ast, \ \forall j$. In such a case, no matter how many single events are detected, one would not be able to decipher the presence of higher-multipoles in the signal without stacking them using the algorithm presented in this work. In such a case, we estimate that one would require to stack $n_0 \simeq 220$ events to reach a nominal detection probability of 95\%.

Note that BBH merger events are detected by search pipelines in coincidence across 2 or more detectors. 
By treating them as independent sources, the number of BBH events may be reduced by factors of $\sim 2$ (double coincident detection) or more! This implies that we may detect higher-multipoles with only 100 events which may be observed in the \texttt{aLIGO} detectors within {a month} of continuous observation at design sensitivity.

\section{Analysis of events in LIGO O1/O2 data}
%

We analysed the events from the O1 and O2 science runs~\cite{GWTC1} for the presence of ${m=3}$ multipoles using 
data from the Gravitational Wave Open Science Center~\cite{GWOSC:catalog}. Some salient points of this analysis are enumerated below:

\begin{enumerate}
\item[a.] 
The parameters of each event was fixed to be the maximum likelihood sample of their respective posterior distributions obtained from parameter estimation studies~\cite{GWTC1PESamples}. 
%
\item[b.] {\emph{Foreground}}: A $5 \,\si{\second}$ segment of strain data chosen $([-4, +1 ]\ \si{\second} $) around the event epoch was taken to be 'on-source' data segment containing the GW signal.
\item[c.] 
{\emph{Background}}: 
LIGO strain data surrounding the event epoch (excluding $\pm 34 \, \si{\second}$ around the coalescence time) were taken as samples of instrumental noise in the detector. 
This data was divided into several non-overlapping segments of $5 \,\si{\second}$ each, whitened by the PSD estimated from longer $64 \,\si{\second}$ segments. The ensemble of noise vectors $N(\alpha)$ obtained from these segments were used to calculate the ensemble mean and noise covariance matrix. The background distribution $p(\beta \, | \, \Hyp_2)$ was evaluated by injecting ${m=2}$ (quadrupole) {\texttt{maxL}} waveforms into each of the $5 \,\si{\second}$ off-source sub-chunks.

\end{enumerate}

Standard data quality vetoes were used to mitigate the effects of problematic data~\cite{Abbott_2018_dq}. An additional veto, analogous to the {gating} technique developed by Usman et al.~\cite{Usman-2015} was used to discard noisy data segments. Under this scheme, any whitened sub-chunk (assumed to be a Gaussian time-series with zero mean and unit variance) having a sample above a nominal gating threshold of $6.0$ was rejected. 

\begin{figure}[t]
\centering
  \includegraphics[width=0.495\textwidth]{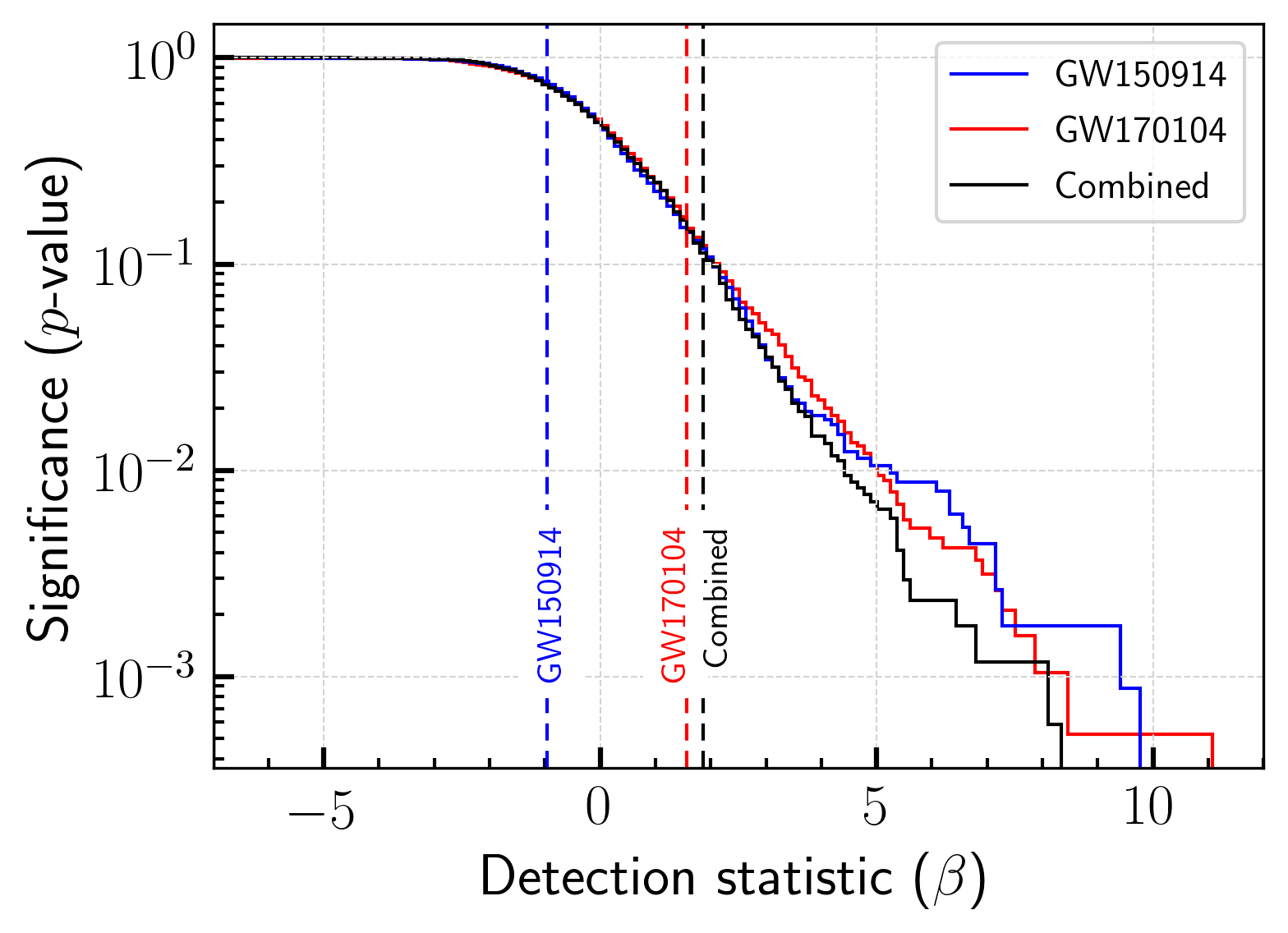}
  \caption[]{
  Analysis of GWTC-1 events: $p$--values determined from the 
  background distribution are shown for two events ({GW150914}, {GW170814}), and after stacking three events ({GW170814}, {GW170818}, {GW170104}). The vertical dashed lines indicate the detection statistic in each case, which can be used to directly infer the significance from corresponding $p$--value traces. Stacking events in the GWTC-1 catalogue leads to a marginal increase of the detection statistic (and improved significance).
   }
  \label{fig:GWTC1}
\end{figure}

In Fig.~\ref{fig:GWTC1}, we show the detection statistic $\beta$ for two events ({GW150914}, {GW170814}) along with the $p$--values calculated from their background distributions. 
GW170104 was found to be the loudest ($\beta = 1.6$, $p$--value = 0.15). 
GW150914, the first and most significant event reported in O1/O2 runs was found with $\beta = -1.0$ ($p$--value = 0.75). 
We also show the combined detection statistic after stacking three most favourable events ({GW170814}, {GW170818}, {GW170104}) in the GWTC-1 catalogue as determined from Eq.~\eqref{eq:selectEvents}. 
%
%
Stacking these events resulted in a marginal increase of the (combined) detection statistic to $\beta = 1.9$ ($p$--value = 0.1), but still far below the nominal detection threshold $\beta^\ast$ at $1\%$ false-alarm probability.

Our analysis is based on a reliable estimation of the ${(2,2)}$ time-frequency track of the events which depends most strongly on the `chirp-mass' of the BBH systems. As the best match-filter template gives a good estimate of the chirp-mass, it may be possible to do prompt follow-up of events for presence of higher-multipoles, immediately after their detection by the search pipelines.

Looking ahead, we would like to follow-up BBH detections made in the recently concluded O3 and upcoming runs of advanced LIGO and Virgo detectors for signatures of $m=3$ multipoles in the signal.
We would also like to extend this framework beyond the inspiral regime to full inspiral-merger-ringdown waveforms, and explore its use in testing general relativity.
With improved detector sensitivities in future, this method could be used to detect other sub-dominant modes of the signal as well. We expect that with increased sensitivity at low frequencies, this method will be very useful for third-generation ground-based detectors (Einstein Telescope~\cite{Punturo_2010}, Cosmic Explorer~\cite{Reitze:2019iox}, Voyager ) to measure the $(2,1)$ mode in particular which will not be possible with advanced LIGO. 
It may also be possible to formulate new tests of GR by demanding consistency of the spacing between time-frequency tracks of different multipoles with theoretical predictions. 

\noindent{\textit{\textbf{Note --}}} While this paper was under revision, the proposed method was employed to search for the presence of higher modes in the gravitational wave events GW190412~\cite{collaboration2020gw190412} and GW190814~\cite{Abbott_2020_gw190814} reported recently by the LVC collaboration and detected the presence of the same.

\vspace{6pt}

\begin{acknowledgements}
This document has a LIGO-DCC No. \texttt{P1900257}. We gratefully acknowledge comments, and feedback from B.~Sathyaprakash, Luc Blanchet, P.~Ajith, M.K.~Das, Ajit~Mehta, Anuradha~Gupta, M.K.~Haris and Nathan Johnson-McDaniel. We thank ICTS Bengaluru for the hospitality, where a part of the manuscript was written.  This work was carried out with the generous funding available from DST's ICPS grant no. {T-150}. S.R. thanks IIT Gandhinagar for SRF. {K.G.A.} acknowledges the Swarnajayanti Fellowship Grant No.DST/SJF/PSA-01/2017-18 of DST-India, Core Research Grant EMR/2016/005594 of SERB, EMR and a grant from the Infosys Foundation. This research has made use of data, software and/or web tools obtained from the Gravitational Wave Open Science Center (\href{https://www.gw-openscience.org}{\texttt{https://www.gw-openscience.org}}), a service of LIGO Laboratory, the LIGO Scientific Collaboration and the Virgo Collaboration.
\end{acknowledgements}


\bibliography{reference}

\clearpage
\appendix
\input{supplement}

\end{document}

%% file: supplement.tex

%

%
%

\twocolumngrid
In this document, we provide detailed calculation of certain crucial results used in the main text. For clarity, we define the notations used elsewhere in this paper.

\begin{table}[htbp!]
\centering
  \begin{tabular}{ l  l  }
  \toprule[1pt]	
  $\bar{x}$     & \ \ CWT of the time-series $x(t)$  \\
  $\hat{x}$     & \ \ Fourier transform of $x(t)$  \\
  $\tilde{x}$   & \ \ Scaleogram of $x(t)$ \\
  $\sim$        & \ \  follows the distribution  \\
  $\dot{\sim}$  & \ \ approximately follows the distribution \\
  $\mathcal{N}(\mu, \sigma^2)$  & \ \ Gaussian distribution with mean $\mu$  \\
                    & \ \ and variance $\sigma^2$ \\
  $\Gamma(a, b)$    & \ \ Gamma distribution with shape-  \\
                    & \ \  parameter $a$ and rate-parameter $b$ \\
  $\vec{\lambda}_j$ & \ \ Parameters of the $j-$th event. \\
  $\beta_j$ ($\beta$) & Single (combined) event detection statistic \\

  \bottomrule[0.5pt]
\end{tabular}
\end{table}  

\section{Continuous wavelet transformation and choice of central frequency of wavelet}
\label{sec:CWT}
The continuous wavelet transformation (CWT) of a signal $x(t)$ in Gabor-Morlet~\cite{Morlet-1984} wavelet basis is given by: \begin{equation}
  \bar{X}(\tau, a) = \frac{1}{\sqrt{a}} \int_{-\infty}^{\infty} x(t)\ \psi^{\ast}\left( \frac{t -\tau}{a}  \right) \: dt, 
  \label{SuppEq:CWT}
\end{equation}
where, $\psi^{\ast}( (t-\tau)/a )$ is the conjugate of the translated and scaled wavelet used. The wavelet is taken to be a square-integrable function parametrised by the scale ($a$) and time translation ($\tau$) parameters. 
%
The energy contained at a specific pixel centered at $(\tau, a)$ is given by the absolute square of $\bar{X}( \tau, a )$;
\begin{equation}
  \tilde{X}(\tau, a) = |   \bar{X}(\tau, a) |^2.
\end{equation}
Analogous to the well-known ``spectrogram'' which represents the energy density of a signal over the time-frequency plane, the scaleogram gives the energy density over the $\tau - a$ parameters.
It can be integrated to extract the total energy of the signal:
\begin{equation}
\label{eq:TimeScaleEnrgy}
  E = \frac{1}{C_g} \int_{-\infty}^{\infty}\int_0^{\infty} \tilde{X}(\tau, a) \: \frac{da}{a^2} d\tau \equiv \norm{x(t)}^2,
\end{equation}
where, $C_g$ is the wavelet admissibility constant satisfying the condition,
\begin{equation}
  C_g = \int_0^{\infty} \frac{ |\hat{\psi}(f)|^2 }{f} df < \infty.
\end{equation}

We used the complex Gabor-Morlet wavelets for the CWT - consisting of a plane wave modulated by a Gaussian envelope:
\begin{equation}
  \psi(\eta; f_0 ) = \frac{1}{\pi^{1/4}} \left( e^{2\pi if_0\eta} - e^{-(2\pi f_0)^2/2} \right) e^{-\eta^2 },
\end{equation}
where $\eta = (t-\tau)/a$. The central frequency of the mother wavelet $f_0$, can also be interpreted as the frequency of the plane wave where $(t-\tau)/a$ is the temporal parameter. In this case, the frequency domain representation of the wavelet has a global maximum at $f_0$. The second term within the bracket is known as the correction term, which preserves the zero mean of the first term, i.e., it corrects for the non-zero mean of the complex plane wave multiplied by the gaussian envelope. In practice, this term can be ignored for $f_0 \gg 0$. In our analysis, for $f_0 > 6/2\pi$, the Gabor-Morlet wavelet can be written in a simpler form as:
\begin{equation}
  \psi(\eta; f_0 ) = \frac{1}{\pi^{1/4}}  e^{2\pi if_0\eta}\: e^{-\eta^2 }
\end{equation}

In order to compare the scaleogram to the spectrogram, we focus on the term of complex plane wave $e^{2\pi if_0\eta}$. The fraction $f_0/a$ can be interpreted as a frequency parameter of the time-frequency representation, and is known as the `pseudo-frequency'. 

Assuming uniform spacing over frequency $f$, the scaleogram calculated using scale parameters $a=f_0/f$ is equivalent to the spectrogram sampled uniformly over time-frequency parameters.
%
%
Thus, the energy contained in a specific time-frequency pixel centred at $(\tau, f)$ is:
\begin{equation}
  \tilde{x}(\tau, f) = \tilde{X} \ \Delta\tau \: \Delta f \equiv \frac{1}{C_g} \tilde{X}(\tau, a) \ \frac{\Delta a}{a^2} \: \Delta \tau, 
\end{equation}
where, $\Delta \tau$ and $\Delta f$ denote the pixel size along time and frequency axes respectively. 
The above definition allows us to interpret the $d$-dimensional template vector $S(\alpha)$ (defined in Eq.~\ref{eq:sigmodel} of the text) to be the vector of signal energy contained in different time-frequency tracks $f(t) = \alpha \, f_{22}(t)$, parameterized by the scaling factor $\alpha$ which takes $d$ discrete value in the interval $[\alpha_{\text{min}}, \; \alpha_{\text{min}}]$. 
 
\vspace{0.5\baselineskip}


\section{Estimation of the noise characteristics}
\label{sec:EstimateNoise}

The noise in the LIGO like detectors is assumed to be approximately stationary and Gaussian with zero mean. 
With this assumption, the noise is fully characterized by the one-sided power spectral density, $S_n(f)$, such that $\Expect[ \hat{n}_{_d}(f) \ \hat{n}_{_d}^{\ast}(f^{'})]  = \frac{1}{2}\delta(f -f^{'})S_{n}(f)$, where $\Expect[ \, \cdot \, ]$ denotes the ensemble average, and $\hat{n}_{_d}(f)$ represents the Fourier transform of the detector output $n_{_d}(t)$. This allows us to produce whitened gaussian noise (WGN) time-series $n(t)$ from the data $n_d(t)$ using the frequency domain relation: $\hat{n}(f) = \hat{n}_{_d}(f)/\sqrt{S_{n}(f)}$. By construction $n(t)$ follows a Gaussian distribution with zero mean and unit variance i.e.  $n(t) \sim \mathcal{N}(0, 1)$.

%
The CWT of a Gaussian time-series follows a complex Gaussian distribution since it is a linear transformation, where both the real and imaginary parts of $\bar{N}(\tau, f)$ follow Gaussian distributions with same variance and zero mean. Further, the spectrogram ($\tilde{N}(\tau, f) = |\bar{N}(\tau, f) |^2$) is the quadrature summation of two Gaussian random variables, and follow a Gamma distribution. 

The noise vectors $N(\alpha)$ are constructed by summing many (typically, several thousands) time-frequency pixels of $\tilde{n}$ (scaleogram of off-source data-segments) along time-frequency trajectories that are scaled with respect to the quadrupole mode trajectory $f_{22}(\tau, \vec \lambda)$. This  implies that the probability distribution of $N(\alpha)$ is a convolution of several thousand Gamma random variables. In this limit, the well-known central limit theorem ensures that $N(\alpha)$ can be approximated by a Gaussian distribution. 
 
 \begin{figure}
\centering
  \includegraphics[width=0.49\textwidth]{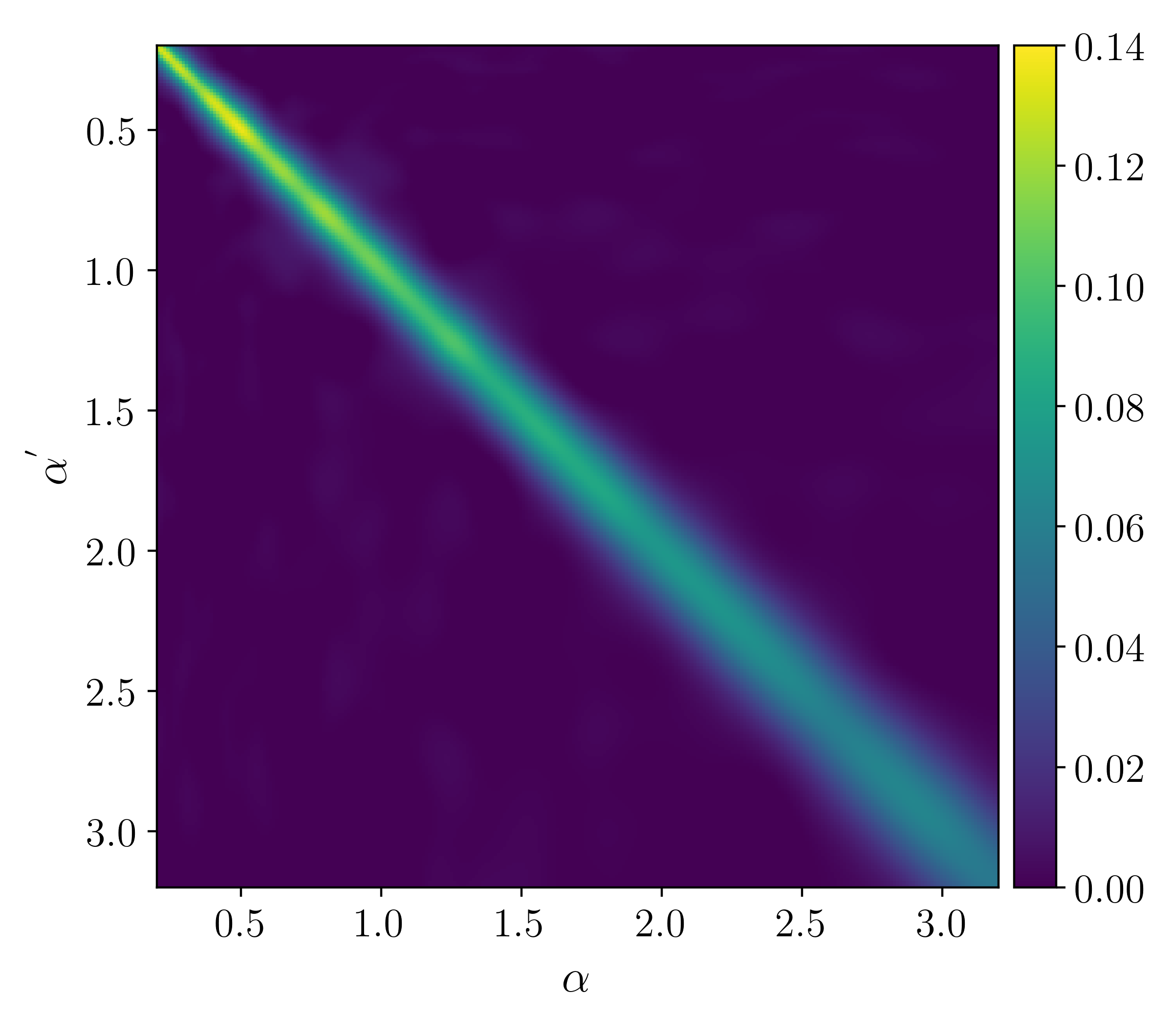}
  \caption{Numerically evaluated noise covariance matrix $\Sigma(\alpha, \alpha')$ for the ensemble of noise vectors $N(\alpha)$.  
  Each noise vector $N(\alpha)$ in the ensemble is calculated from the scaleogram of a distinct realization of synthetic \texttt{aLIGO} noise. The scaleogram pixels are summed along time-frequency tracks scaled with respect to a fiducial quadrupole-mode trajectory of a  BBH system with component masses $[35, 6]\msun$ and effective spin $-0.3$. 
 }
  \label{fig:cov_mat}
\end{figure}

Not only are the scaleogram pixels along a track correlated with each other, but the summation of pixels along two nearby tracks are also highly correlated. We can characterize this correlation by numerically evaluating the covariance matrix $\Sigma(\alpha, \alpha')$ from an ensemble  of many $N(\alpha)$ vectors (one from each off-source segment):
\begin{equation}
\label{eq:NumericalCovariance}
  \Sigma(\alpha, \alpha') \coloneq \mathbb{E}\left [ \left(N(\alpha) - \mu(\alpha)\right) \,  \left(N(\alpha') - \mu(\alpha')\right)^T \right ], 
  \end{equation}
where $\mu(\alpha) = \mathbb{E}\left [ N(\alpha) \right ]$ is the ensemble average. 

An example of a numerically estimated covariance matrix (for synthetic \texttt{aLIGO} data) is shown in Fig.~\ref{fig:cov_mat}. 
It is seen that the covariance matrix is non-diagonal, especially the off-diagonal elements close to the principal diagonal are comparable to the values of the main diagonal elements.


\section{Details of hypothesis testing for the composite signal model}
\label{sec:SupHypothesisTesting}

Here we discuss the details of the hypothesis testing in additive correlated Gaussian noise for detecting the sub-dominant modes of a single BBH merger event. 

As defined in Eq.~\ref{eq:sigmodel}, the template vector $S_j(\alpha)$ is calculated from $\tilde{s}_j(\tau, f)$; where the scaling parameter $\alpha$ takes by varying the scaling parameter $d$-discrete steps between $\alpha_\text{min} \leq \alpha \leq \alpha_\text{max}$. 
As such, $S(\alpha) \equiv \vec S$ can be considered to be a vector in a $d$-dimensional Euclidean vector space $\mathbb{R}^d$. Similarly, the noise vectors $N(\alpha)$ constructed from off-source data surrounding the event,  and the observational data vector $Y(\alpha)$ constructed from the on-source data segment containing the event epoch, can also be treated as vectors in $\mathbb{R}^d$. 

Let $\vec{N} \sim \mathcal{N}(\vec{\mu}, \, \Sigma_{d \times d})$ be a correlated Gaussian random vector in a $d-$dimensional vector space. For simplicity, we first consider a binary hypotheses:  the null hypothesis $\Hyp_0$, that the observed data $\vec{Y}$ is due to instrumental noise $\vec{N}$ only; and its alternative $\Hyp_1$, that $\vec{Y}$ is due to a signal embedded in noise, i.e. $\vec{N}+\vec{S}$. 
The likelihood of $\vec{Y}$ under the two hypotheses are given by: 
\begin{equation}
  \begin{split}
  p(\vec{Y} \mid \Hyp_0) & = \frac{\exp \left[ -\frac{1}{2} (\vec{Y} - \vec{\mu})^T\inv{\Sigma}\vec{Y} \right]}{\sqrt{(2\pi)^d |\inv{\Sigma}|} },   \\
  p(\vec{Y} \mid \Hyp_1) & = \frac{\exp \left[-\frac{1}{2} (\vec{Y}-\vec{\mu}-\vec{S})^T \inv{\Sigma}(\vec{Y}-\vec{\mu}-\vec{S}) \right]}{\sqrt{(2\pi)^d |\inv{\Sigma}|}}, 
  \end{split}
\end{equation}
where $\vec \mu$ is the ensemble average of the noise vectors and $\Sigma(\alpha, \alpha')$ is the noise covariance matrix.  $|\inv{\Sigma}|$ denotes the determinant of $\inv{\Sigma}$. The \emph{logarithmic likelihood ratio} is given by
\begin{equation}
\label{eq:LLR1}
  \Lambda = (\vec{Y} - \vec{\mu})^T\inv{\Sigma}\vec{S} - \frac{1}{2}\vec{S}^T\inv{\Sigma}\vec{S} .
\end{equation}

If the null hypothesis $\Hyp_0$ is true, then one can show that $\Lambda \sim \mathcal{N}( -\gamma^2/2, \, \gamma^2 )$~\cite{gallager_2013}. On the other hand, $\Lambda \sim \mathcal{N}(\gamma^2/2, \, \gamma^2 )$ when $\Hyp_1$ is true. $\gamma$ is the norm of the signal embedded in noise, i.e. $\gamma^2 = \vec{S}\, \inv{\Sigma}\, \vec{S}$.
Motivated by these results, we define a new detection statistic,
\begin{align}
  \label{eq:betaStatistic0Sup}
  \beta &= \left(\Lambda + \gamma^2/2 \right)/\gamma,  \\
        &= \langle \vec{Y} - \vec{\mu} \, \mid \, \vec{S} \rangle/\gamma,
\end{align}
which follows $p(\beta) \sim \mathcal{N}(0, 1)$ under $\Hyp_0$. As expected this result is independent of the signal parameters. 
On the other hand, $\beta$ follows $\mathcal{N}(\gamma, 1)$ when $\Hyp_1$ is true. Here, $\langle \cdot \vert \cdot \rangle$ denotes the inner-product between two vectors inversely weighted by the covariance matrix $\Sigma(\alpha, \alpha')$.


%

\begin{figure}
\centering
  \includegraphics[width=0.49\textwidth]{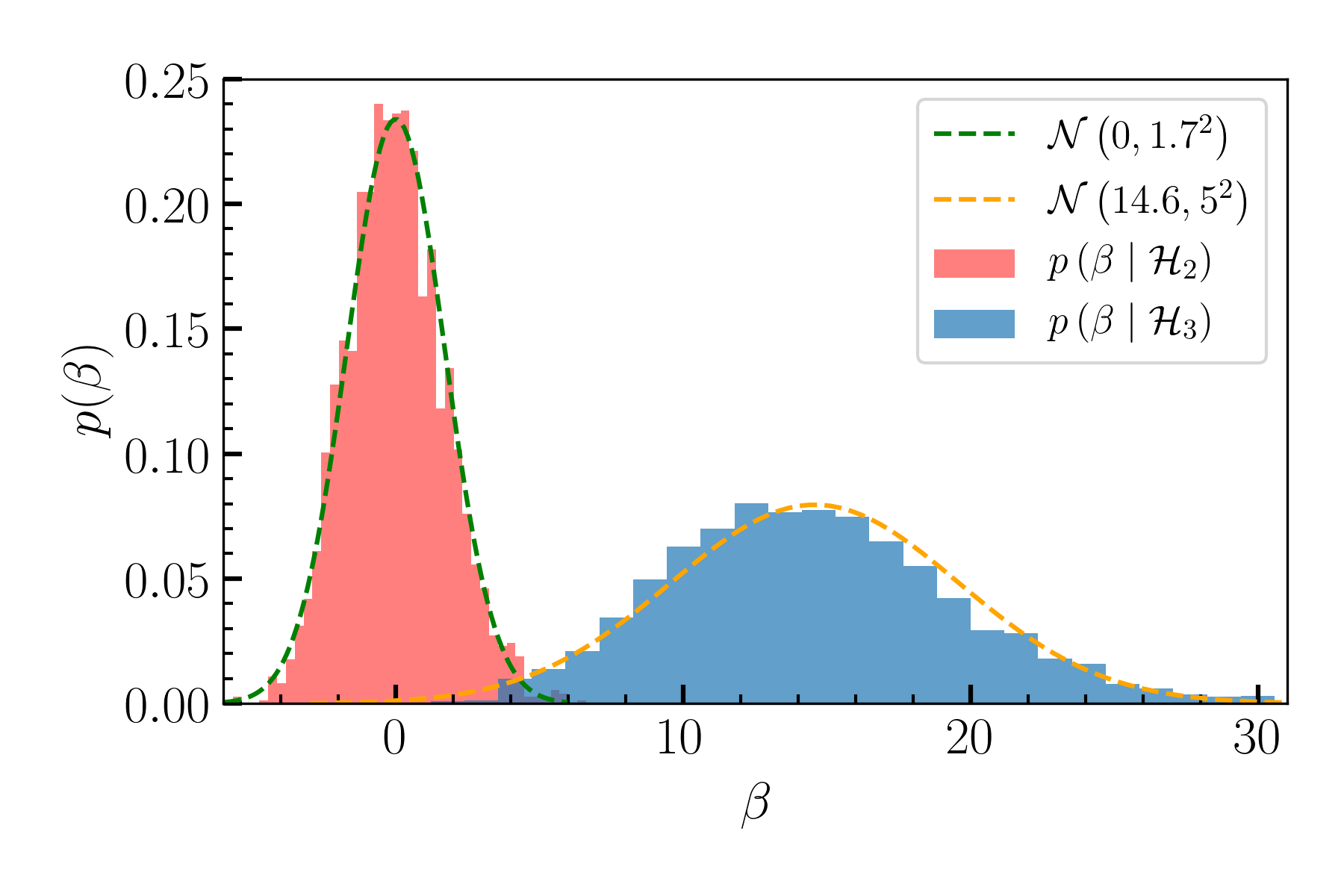}
  \caption{Distribution of the detection statistic $\beta$ for two cases: $\mathcal{H}_2$ true, i.e. when data contains only the dominant quadrupole component of the signal (\emph{red-histogram}, to the left) and $\mathcal{H}_3$ true, when the next higher $(m=3)$ multipoles are also preesent in the signal (\emph{blue-histogram}, to the right).
  The background distribution $p(\beta \mid \mathcal{H}_2)$ is shown to agree with a Gaussian distribution $\mathcal{N}(0, 1.7^2)$. In presence of higher-multipoles of the signal, the distribution $p(\beta \mid \mathcal{H}_3)$ is shown to agree with a Gaussian distribution with mean equal to the optimum signal norm $\gamma_3^{\text{inj}}$. Results shown are obtained from BBH signal injections with component masses $[32, 6]\msun$, effective spin $-0.3$, inclination angle $97^{\circ}$ and fixed SNR of $36.7$.
  }
  \label{fig:dist_sngl_evn}
\end{figure}

In the main section of the paper, we have defined not two but three composite hypotheses as given in Eq.~\ref{eq:DefHypotheses}. 
The LLR $\Lambda_3(a_2, a_3)$, which quantifies the odds of observing $Y(\alpha)$ under $\Hyp_3$ to 
that under the null hypothesis $\Hyp_0$ is: 
\begin{equation}
\begin{split}
  \Lambda_3(a_2, a_3) & = \langle Y(\alpha) - \mu(\alpha) \mid a_2 S_2(\alpha) + a_2 S_2(\alpha) \rangle - \\
  & \frac{1}{2} \langle a_3S_3(\alpha)  \mid a_3S_3(\alpha) \rangle  + \frac{1}{2} \langle a_2S_2(\alpha)  \mid a_2S_2(\alpha) \rangle,
  \end{split}
\end{equation}
where the mutually independent $S_2(\alpha)$ and $S_3(\alpha)$ template vectors, corresponding to the $m=2$ and $m=3$ signal multipoles respectively, are each defined upto a free overall amplitude parameter. These parameters $a_2$ and $a_3$ are fixed by maximising the above LLR, i.e.
\begin{equation}
\label{eq:AmplitudefactorSup}
a_{2, 3}^{\ast} = \argmax_{a_2,\, a_3}\, \Lambda_3(a_2, a_3).
\end{equation} 
%
%
Note that $\Lambda_3(a_2, a_3=0)$ is identical to $\Lambda_2(a_2)$ where the latter is the LLR of observing $Y(\alpha)$ under $\Hyp_2$ as compared to $\Hyp_0$,

The difference between the maximum likelihood values $\Lambda_3(a_2^{\ast}, a_3^{\ast})$ and $\Lambda_2(a_2^{\ast})$ indicate which of the two competing hypotheses $\Hyp_3$ and $\Hyp_2$ is favoured by the data. 

Motivated by the detection statistic defined earlier in Eq.~\eqref{eq:betaStatistic0Sup} for the case of a binary hypotheses, we can write its equivalent for our present case to test if the residual $Y(\alpha) - a_2^{\ast}\, S_2(\alpha)$ contains the higher-multipole ($m=3$) signal embedded in noise: 
\begin{equation}
\label{eq:DetecticionStatisticSup}
  \beta = \langle Y(\alpha) - \mu(\alpha) -  a_2^{\ast}\, S_2(\alpha) \, \mid \, a_3^{\ast}\, S_3(\alpha) \rangle\, / \; \gamma_3,
\end{equation}
where, $\gamma_3 = \lVert a_3^{\ast} \, S_3(\alpha)\rVert$ is the template norm. 

By evaluating $\beta$ repeatedly after injecting only the dominant quadrupole mode of a signal (with fixed set of parameters) in many different noise realisations, $\beta$ can be shown to follow a zero-mean, normal distribution: $p(\beta \mid \Hyp_2) \sim \mathcal{N}(0, \, \text{var}>1)$. This serves as the background distribution against which the significance of the results are evaluated.

On the other hand, when a signal (with fixed parameters) containing the dominant {\emph{and}} the next-higher (m=3) harmonic with norm $\gamma^{\text{inj}}_3$ is injected in several different noise realizations, then $\beta$ can be shown to be distributed as: $p(\beta \mid \Hyp_3) \sim \mathcal{N}(\gamma^{\text{inj}}_3, \, {\text{var}}>1)$; with the mean of the distribution being $\langle \beta \rangle = \gamma^{\text{inj}}_3$.

The variance of the distributions are greater than unity due to the cross-terms between the injected signal and noise in the  spectrogram, and is inherent to the method presented in this work. 

%


In Fig.~\ref{fig:dist_sngl_evn}, we show the distribution of the detection statistic $\beta$ from a simulation where 
\begin{itemize}
\item[a.]  
	at first, only the dominant $(m=2)$ multipole of the GW signal from a BBH system was injected in $2500$ realisations
	of synthetic \texttt{aLIGO} noise. This corresponds to the hypothesis $\Hyp_2$ in Eq.~\ref{eq:DefHypotheses} to be true. 
\item[b.] 
	next, the next-higher $(m=3)$ multipole of the signal was also included during injections, which corresponds to hypothesis $
	\Hyp_3$ in Eq.~\ref{eq:DefHypotheses} to be true.
\end{itemize}
The component masses of the BBH system were chosen to be $[32, 6]\msun$, with effective spin $-0.3$. The  orientation of the binary was fixed to be $97^{\circ}$ with respect to the line of sight. 

The resulting distributions $p(\beta \mid \Hyp_2)$ and $p(\beta \mid \Hyp_3)$ of the detection statistic, for the two cases are shown in {Fig.~\ref{fig:dist_sngl_evn}}. In the presence of only the quadrupole mode of the signal in noise, the detection statistic has a $\mathcal{N}(0, 1.7^2)$ distribution with zero mean. On the other hand, when higher-multipoles are included, the the distribution shifts to the right such that the mean value $\langle \beta \rangle $ is equal to the optimal signal norm $\gamma_3^{\rm{opt}}$ of the $m=3$ template. 

The width of the background distribution $p\left(\beta \mid \Hyp_2\right)$ depends on the signal power contributed by the dominant quadrupole mode of the signal (in the limit of no power contributed from the $m=2$ mode, this width becomes $1$). We can estimate this width numerically by injecting quadrupole waveforms in a set of noise realisations.

As the background distributions of each of the event is different, it poses a challenge while comparing the detection statistic $\beta$ across multiple events. For making comparisons, it is prudent to scale the detection statistic ($\beta$) of each event  by the corresponding standard deviation of $p\left(\beta \mid \Hyp_2\right)$. By such a scaling, the background distributions of all the events are effectively reduced to $\mathcal{N}(0, 1)$, thereby making meaningful comparisons possible.




The nominal threshold of detection $\beta^\ast$ is set at a value corresponding to 1\% false-alarm probability, which is obtained by numerically solving the equation $\int_{\beta^{\ast}}^{\infty} p\left(\beta \mid \Hyp_2\right) \, d\beta = 0.01$ for  $\beta^\ast$.